\documentclass[twocolumn,showpacs,aps,prl,superscriptaddress,letterpaper]{revtex4}
\usepackage{graphicx}
\usepackage{dcolumn}
\usepackage{amsmath}
\usepackage{epsfig}
\long\def\inst#1{\par\nobreak\kern 4pt\nobreak
    {\itshape #1}\par\vskip 10pt plus 3pt minus 3pt}
\RequirePackage{xspace}
\usepackage{relsize}
\def\Bztorhozrhoz {\ensuremath{\Bz \to \rho^0 \rho^0 }\xspace}
\def\Btozz {\ensuremath{\Bz \to \rho^0 \rho^0 }\xspace}
\def\rhozrhoz {\ensuremath{\rho^0\rho^0 }\xspace}
\def\Bztorhoprhom {\ensuremath{\Bz \to \rho^+ \rho^- }\xspace}
\def\Bptorhozrrhop {\ensuremath{\Bp \to \rho^+ \rho^0 }\xspace}

\def\babar{\mbox{\slshape B\kern-0.1em{\smaller A}\kern-0.1em
    B\kern-0.1em{\smaller A\kern-0.2em R}}}
\def\Dbar    {\kern 0.18em\overline{\kern -0.18em D}{}\xspace}
\def\Bbar    {\kern 0.18em\overline{\kern -0.18em B}{}\xspace}

\def\BB      {\ensuremath{B\Bbar}\xspace} 
\def\Bz      {\ensuremath{B^0}\xspace}
\def\Bzb     {\ensuremath{\Bbar^0}\xspace}
\def\BzBzb   {\ensuremath{\Bz {\kern -0.16em \Bzb}}\xspace}
\def\Bu      {\ensuremath{B^+}\xspace}
\def\Bub     {\ensuremath{B^-}\xspace}
\def\Bp      {\ensuremath{\Bu}\xspace}

\def\BpBm    {\ensuremath{\Bu {\kern -0.16em \Bub}}\xspace}

\newcommand{\optbar}[1]{\shortstack{{\tiny (\rule[.4ex]{1em}{.1mm})}
  \\ [-.7ex] $#1$}}
\def\BorBbar    {\kern 0.18em\optbar{\kern -0.18em B}{}\xspace}
\def\DorDbar    {\kern 0.18em\optbar{\kern -0.18em D}{}\xspace}
\def\KorKbar    {\kern 0.18em\optbar{\kern -0.18em K}{}\xspace}
\def\CP                {\ensuremath{C\!P}\xspace}
\def\pep2{PEP-II}
\mathchardef\Upsilon="7107
\def\Y#1S{\ensuremath{\Upsilon{(#1S)}}\xspace}

\def\FourS {\Y4S}

\def\CP                {\ensuremath{C\!P}\xspace}

\def\u     {\ensuremath{u}\xspace}

\def\d     {\ensuremath{d}\xspace}

\def\b     {\ensuremath{b}\xspace}

\def\pip   {\ensuremath{\pi^+}\xspace}
\def\pim   {\ensuremath{\pi^-}\xspace}

\def\upsbb   {\ensuremath{\FourS \to \BB}\xspace}

\def\invfb   {\ensuremath{\mbox{\,fb}^{-1}}\xspace}
\newcommand{\mev}{\ensuremath{\mathrm{\,Me\kern -0.1em V}}\xspace}
\def\B       {\ensuremath{B}\xspace}
\def\mes        {\mbox{$m_{\rm ES}$}\xspace}
\def\DeltaE     {\mbox{$\Delta E$}\xspace}
\def\Dm      {\ensuremath{D^-}\xspace}

\newcommand{\jprlBase}       {Phys.\ Rev.\ Lett.\xspace}
\newcommand{\jprl}      [1]  {\jprlBase\ {\bf #1}}

\newcommand{\BABARPubYear}     {04}
\newcommand{\BABARPubNumber}  {048}

\newcommand{\SLACPubNumber} {10909}

\begin{document}

\begin{flushleft}
\babar-PUB-\BABARPubYear/\BABARPubNumber\\
SLAC-PUB-\SLACPubNumber\\[10mm]
\end{flushleft}

\title{
\large \bfseries \boldmath
Limit on the \Btozz Branching Fraction and Implications for the
CKM Angle $\alpha$
}

\author{B.~Aubert}
\author{R.~Barate}
\author{D.~Boutigny}
\author{F.~Couderc}
\author{Y.~Karyotakis}
\author{J.~P.~Lees}
\author{V.~Poireau}
\author{V.~Tisserand}
\author{A.~Zghiche}
\affiliation{Laboratoire de Physique des Particules, F-74941 Annecy-le-Vieux, France }
\author{E.~Grauges-Pous}
\affiliation{IFAE, Universitat Autonoma de Barcelona, E-08193 Bellaterra, Barcelona, Spain }
% ...
\collaboration{The \babar\ Collaboration}
\noaffiliation


\date{December 23, 2004}

\begin{abstract}
We search for the decay $B^0\rightarrow \rho^0\rho^0$ in a data 
sample of about 227 million 
$\Upsilon (4S)\rightarrow B\kern 0.18em\overline{\kern -0.18em B}$
decays collected with the $\babar$ detector at the
PEP-II asymmetric-energy $e^+e^-$ collider at SLAC.  
We find no significant signal and set an upper limit of 
$1.1\times  10^{-6}$ at 90\% CL on the branching fraction.
As a result, the uncertainty due to penguin contributions
on the CKM 
unitarity angle $\alpha$ measured in $B\to\rho\rho$ 
decays is 11$^{\mathrm o}$ at 68\% CL.
\end{abstract}


\maketitle


Measurements of \CP-violating asymmetries in the \BzBzb system provide
tests of the standard model by over-constraining the 
Cabibbo-Kobayashi-Maskawa (CKM)
quark-mixing matrix~\cite{CabibboKobayashi}
through the measurement of the unitarity angles. 
Measuring the time-dependent \CP asymmetry in a neutral-$B$-meson 
decay to a \CP eigenstate dominated by the tree-level amplitude 
$\b \to \u{\bar\u}\d$
gives an approximation $\alpha_{\rm eff}$ to the CKM unitarity angle 
$\alpha\equiv \arg\left[-V_{td}^{}V_{tb}^{*}/V_{ud}^{}V_{ub}^{*}\right]$.
The correction $\Delta\alpha= \alpha-\alpha_{\rm eff}$,
which accounts for the effects of penguin-amplitude contributions 
as an additional decay mechanism, 
can be extracted from an isospin analysis of the 
branching fractions of the $B$ decays into the full set of 
isospin-related channels~\cite{gronau90}. 

Measurements of branching fractions and time-dependent \CP 
asymmetries in $B\to\pi\pi$, $\rho\pi$, and $\rho\rho$
have already provided information on $\alpha$. 
Because the branching fraction for $B^0\to\pi^0\pi^0$ is
comparable to that for 
$B^+\to\pi^+\pi^0$ and $B^0\to\pi^+\pi^-$,
the limit on the correction is weak:
$|\Delta\alpha_{\pi\pi}| < 35^{\mathrm o}$ 
at 90\% confidence level (CL)~\cite{pi0pi0}. 
(Charge conjugate $B$ decay modes are implied in this paper.)
In contrast, the $\rho^0\rho^0$ channel has a much 
smaller branching fraction than the channels with charged 
$\rho$'s~\cite{vvbabar, rho0rhopbelle, rhoprhom, rhoprhom2}.  
As a consequence, it is possible to
set a tighter limit on $\Delta\alpha_{\rho\rho}$.
This makes the $\rho\rho$ system particularly effective for 
measuring $\alpha$ in a model-independent way. 

In $B\to\rho\rho$ decays the final state is 
a superposition of \CP-odd and \CP-even states,
and an isospin-triangle relation~\cite{gronau90} holds for each 
of the three helicity amplitudes, which can be separated through 
an angular analysis. The measured polarizations in 
\Bptorhozrrhop~\cite{vvbabar,rho0rhopbelle} 
and \Bztorhoprhom~\cite{rhoprhom, rhoprhom2} modes 
indicate that the $\rho$'s are nearly entirely 
longitudinally polarized.
The current best limit on the $B^0\to\rhozrhoz$ branching 
fraction was obtained by \babar\ with a
sample of 89 million \upsbb decays~\cite{vvbabar}. 

\begin{figure}[htbp]
\setlength{\epsfxsize}{1.0\linewidth}\leavevmode\epsfbox{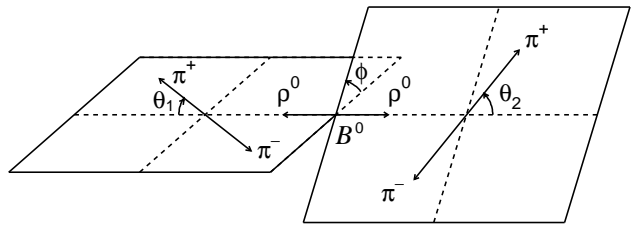}
\caption{Definition of helicity angles $\theta_1$, $\theta_2$, and
$\phi$ for the decay ${B^0\to\rho^0\rho^0}$. The $\rho^0$
final states are shown in the $\rho^0$ rest frames.}
\label{fig:helangles}
\end{figure}


In this Letter we present improved constraints on the \Btozz branching 
fraction and the penguin contribution to the measurement of the 
unitarity angle $\alpha$.
These results are based on data collected
with the \babar\ detector~\cite{babar} at the PEP-II asymmetric-energy
$e^+e^-$ collider~\cite{pep2} located at the Stanford Linear Accelerator
Center. A sample of $226.6\pm 2.5$ million $\BB$ pairs,
corresponding to an integrated luminosity of approximately 205~\invfb, 
was recorded at the $\FourS$ resonance with the center-of-mass (c.m.) energy 
$\sqrt{s} = 10.58$ GeV.
We use a sample of 16~\invfb taken 40~\mev below the $\FourS$
resonance to study background contributions from 
$e^+e^-\rightarrow q\bar{q}~( q = u, d, s, \mathrm{or}~c)$ 
continuum events.


To reconstruct $\Bztorhozrhoz\to(\pi^+\pi^-)(\pi^+\pi^-)$ 
candidates, we select four charged tracks that 
are consistent with originating from a single vertex near 
the $e^+e^-$ interaction point.
Particle identification is provided by 
measurements of the energy loss in
the silicon vertex tracker and the drift chamber
and by the Cherenkov angle in
an internally reflecting ring-imaging Cherenkov detector.

The angular distribution of the $\Bztorhozrhoz$ decay products
can be expressed as a function of the helicity angles 
($\theta_1$, $\theta_2$, ${\phi}$), which are
defined by the directions of the two-body $\rho^0$ decay 
axes and the direction opposite the \B in the $\rho^0$ 
rest systems, as shown in Fig.~\ref{fig:helangles}.
Since the detector acceptance does not depend on $\phi$, 
the resulting angular distribution 
${d^2\Gamma / (\Gamma\,d\!\cos \theta_1\,d\!\cos \theta_2)}$ is
\begin{eqnarray}
{9 \over 4} \left \{ {1 \over 4} (1 - f_L)
\sin^2 \theta_1 \sin^2 \theta_2 + f_L \cos^2 \theta_1 \cos^2 \theta_2 \right\},
\label{eq:helicityshort}
\end{eqnarray}
\noindent where $f_L=|A_0|^2/(\Sigma|A_\lambda|^2)$ is the 
longitudinal polarization fraction and
$A_{\lambda=-1,0,+1}$ are the helicity amplitudes.

The identification of signal $B$ candidates is based 
on two kinematic variables: 
the beam-energy-substituted mass,
$\mes = [(s/2 + {\mathbf {p}}_i\cdot {\mathbf{p}}_B)^2/E_i^2-
{\mathbf {p}}_B^2]^{1/2}$,
where the initial total $e^+e^-$
four-momentum $(E_i, {\mathbf {p_i}})$ and the \B
momentum ${\mathbf {p_B}}$ are defined in the laboratory frame; 
and the difference between the reconstructed \B energy in the
c.m. frame and its known value
$\DeltaE = E_B^{\rm cm} - \sqrt{s}/2$.
The signal $m_{\rm{ES}}$ and $\DeltaE$ resolutions are
2.6 MeV/$c^2$ and 20~\mev, respectively.
The selection requirements for $m_{\rm{ES}}$, $\DeltaE$,
the two $\pi^+\pi^-$ invariant masses $m_{1,2}$, 
and the helicity angles are the following:
$5.24 < \mes < 5.29$ GeV/$c^2$, $|\DeltaE|<85~\mev$,
$0.55 < m_{1,2} < 1.00$ GeV/$c^2$,  
and $|\cos\theta_{1,2}|<0.99$. 
The latter requirement removes a region with low reconstruction 
efficiency.

To reject the dominant continuum 
background we require $|\cos\theta_T| < 0.8$, where $\theta_T$ 
is the angle between the $B$-candidate thrust axis
and that of the remaining tracks and neutral clusters in
the event, calculated in the c.m. frame. 
Other discriminating variables include the polar 
angles of the $B$ momentum vector and the $B$-candidate thrust 
axis with respect to the beam axis in the c.m. frame, 
and the two Legendre moments $L_0$ and $L_2$ of the energy 
flow around the $B$-candidate thrust axis~\cite{bigPRD}.
These variables are combined in a neural network, the output of 
which is transformed into a variable ${\cal E}$ for which the 
signal and background distributions are approximately Gaussian.

We veto the background mode $\Bz\to\Dm\pip\to h^+\pim\pim\pip$, 
where $h^+$ refers to a pion or kaon. We require the 
invariant mass of the three-particle combination
that excludes the highest-momentum track in the 
candidate \B rest frame to be inconsistent with being 
the $D$-meson mass ($|m_{h\pi\pi}-m_D| >$ 13 MeV/$c^2$).
After application of all selection criteria, 
$N_{\rm cand}=35740$ events are retained, 
most of which are background events  
with candidates in the sidebands of the variables.
On average each selected event has 1.05 candidates.
When more than one candidate is present in the same event, 
one candidate is selected randomly.

The signal selection efficiency determined
from Monte Carlo (MC)~\cite{GEANT} simulation
is 27\% or 32\% for longitudinally or transversely 
polarized events, respectively.
MC simulation shows that 22\% of longitudinally
and 8\% of transversely polarized signal 
events are misreconstructed with one or more tracks
not originating from the $B^0\to\rho^0\rho^0$ decay.
These are mostly due to combinatorial background from 
low-momentum tracks from the other \B.
We treat these as part of the signal.

Further background separation is achieved by
the use of multivariate $B$-flavor-tagging
algorithms trained to
identify primary leptons, kaons, soft pions and
high-momentum charged particles
from the other $B$ in the event~\cite{babarsin2beta}.
The discrimination power arises from the difference between 
the tagging efficiencies for signal and background in the five 
tagging categories $c_{\rm tag}$.


We use an unbinned extended maximum likelihood fit to extract
the $B^0\to\rho^0\rho^0$ event yield. The likelihood function is
\begin{equation}
{\cal L} = \exp\left(-\sum_{k}^{} n_{k}\right)\, 
\prod_{i=1}^{N_{\rm cand}} 
\left(\sum_{j}~n_{j}\, 
{\cal P}_{j}(\vec{x}_{i})\right),
\label{eq:likel}
\end{equation}
where $n_j$ is the number of events for each hypothesis $j$
(signal, continuum, and six $B$-background classes), and 
${\cal P}_{j}(\vec{x}_{i})$ is the corresponding 
probability density function (PDF), evaluated with 
the variables 
$\vec{x}_{i}=\{m_{\rm{ES}}, \Delta E, {\cal E}, 
m_1, m_2, \cos\theta_1, \cos\theta_2, c_{\rm tag}\}$
of the $i$th event.

We use MC-simulated events to study the background from 
other $B$ decays. The charmless modes are grouped into five 
classes with similar kinematic and topological properties: 
$B^0\to a_1^{\pm}\pi^{\mp}$; $B^0\to \rho^0K^{*0}$;
$B^+\to\rho^+\rho^0$; a combination of $B\to\rho\pi$ and 
$B^0\to\rho^+\rho^-$; and $B$ decays to other charmless modes 
not included explicitly.
One additional class accounts for the remaining neutral and 
charged $B$ decays to charm modes. The number of events in each 
class $n_{j}$ is left free in the fit with the exception of 
three classes where $n_{j}$ is fixed 
either to the expectations from independent measurements
($78\pm20$ events of $B^+\to\rho^+\rho^0$, 
and $48\pm 8$ events of $B\to\rho\pi$ and $B^0\to\rho^+\rho^-$) 
or to the extrapolation from the flavor-SU(3)-related
$B$-decay modes~\cite{vvbabar}
($25\pm 18$ events of $B^0\to \rho^0K^{*0}$).


Since the correlations among the variables are found to be small, 
we take each ${\cal P}_j$ as the product of the PDFs for the 
separate variables. Exceptions are the correlation between the two 
helicity angles in signal, and mass-helicity correlations in
backgrounds and misrecontructed signal, taken into account as 
discussed below.

We use double-Gaussian functions to parameterize the 
$m_{\rm{ES}}$ and $\Delta E$ PDFs for signal,
and a relativistic $P$-wave Breit-Wigner (BW) 
convoluted with a Gaussian resolution function
for the resonance masses.
The angular distribution for signal, expressed as a function of the 
longitudinal polarization in Eq.~(\ref{eq:helicityshort}), is multiplied 
by a detector acceptance function ${\cal G}(\cos\theta_1, \cos\theta_2)$,
obtained with MC simulation. 
The distributions of misreconstructed signal events 
are parameterized with empirical shapes in a fashion similar 
to that used for $B$ background, as described below.
The ${\cal E}$ variable is described by two asymmetric 
Gaussian functions with different parameters for signal
and background distributions.

The $m_{\rm{ES}}$ distribution of the continuum background
is described with the ARGUS parameterization~\cite{argus}.
The $\Delta E$ and resonance mass $m_{1,2}$ 
PDFs are parameterized with low-degree polynomials. 
The parameterization of the $m_1$ and $m_2$ distributions
includes a BW resonant component to account for
the real $\rho^0$ resonances in the continuum background,
which are assumed to be unpolarized 
and thus to have a flat distribution in $\cos\theta_{1,2}$.
The $\cos\theta_{1,2}$ distribution of the continuum 
background excluding the real resonances is parameterized 
with a second-degree polynomial and an exponential function 
to allow for the increased fraction of 
combinatorial $\pi^+\pi^-$ candidates with low momentum 
pions near $|\cos\theta_{1,2}|=1$.
This parameterization depends on the $\rho$ candidate's mass.

The PDFs for exclusive non-signal \B decay modes are 
generally modeled with empirical non-parametric 
distributions~\cite{nonparam}.
However, analytical distributions are used for
the variables that have distributions 
identical to those for signal, such as $m_{\rm{ES}}$
when all four tracks come from the same $B$, or $\pi^+\pi^-$
invariant mass $m_{1,2}$ when both tracks come from 
a $\rho^0$ meson.
The two $\rho^0$ candidates of some exclusive non-signal modes 
can have very different mass and helicity distributions.
This occurs when one of the two $\rho^0$ candidates 
is real (e.g., $\rho^+\rho^0$, $\rho^0K^{*0}$) 
or when one of the two $\rho^0$ candidates contains a 
high-momentum pion ($a_1\pi$). In such cases, we use a 
four-variable correlated mass-helicity PDF.

The signal and $B$-background PDF parameters are extracted from 
MC simulation while the continuum background PDF parameters 
are obtained from data in $m_{\rm{ES}}$ and $\Delta E$ sidebands. 
The MC parameters of 
$m_{\rm{ES}}$, $\Delta E$, and ${\cal E}$ are adjusted by 
comparing data and MC in calibration channels with similar 
kinematics and topology,
such as $B^0\to D^-\pi^+$ with $D^-\to K^+\pi^-\pi^-$.
Finally, the $B$-flavor tagging PDFs for all categories are
the discrete $c_{\rm tag}$ distributions of tagging efficiencies.
Large samples of fully reconstructed $B$-meson decays are
used to obtain the $B$-tagging efficiencies for signal $B$ decays 
and to study systematic uncertainties in the MC values 
of $B$-tagging efficiencies for the $B$ backgrounds.

\begin{figure}[t]
\centerline{
\setlength{\epsfxsize}{0.5\linewidth}\leavevmode\epsfbox{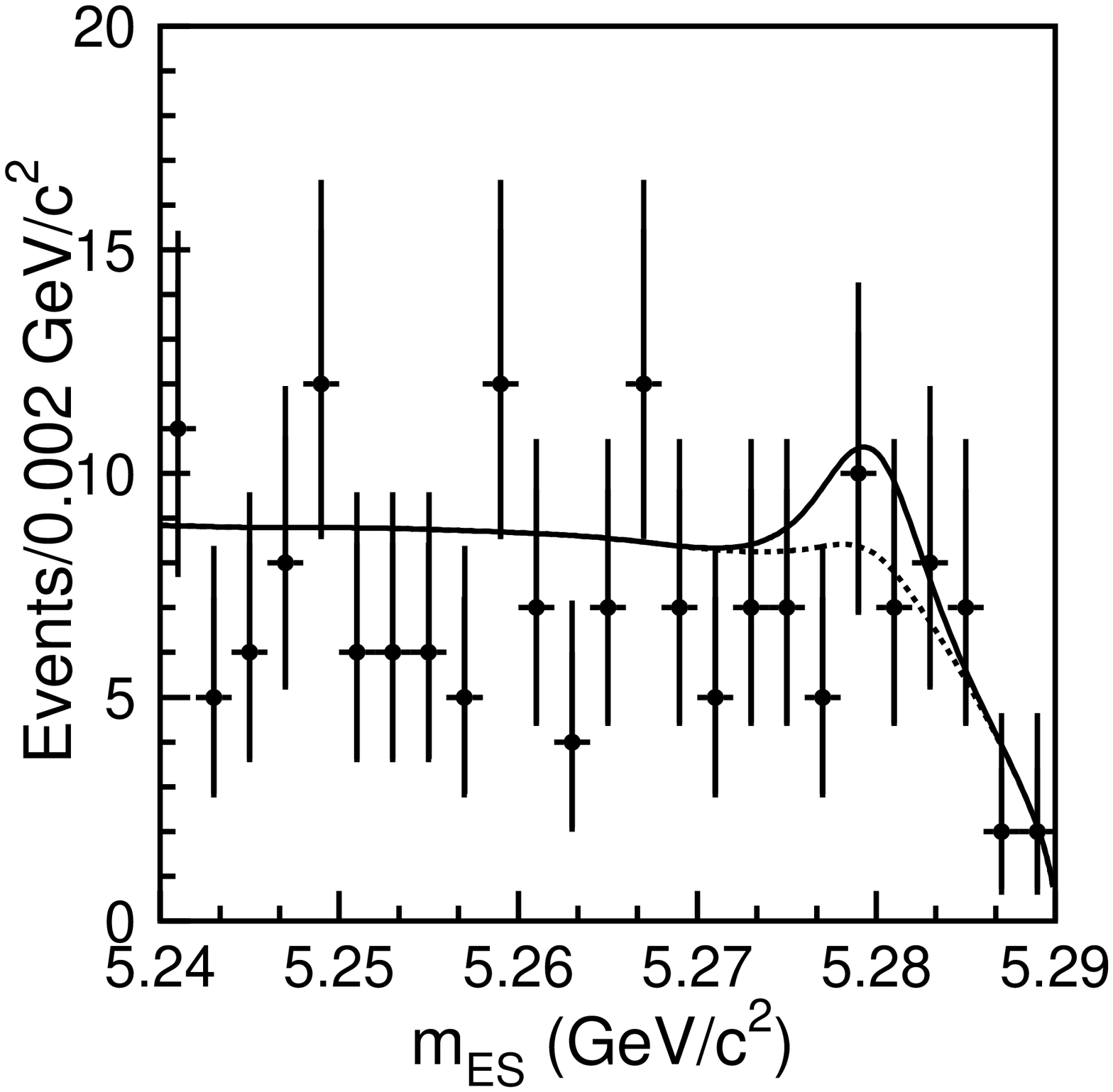}
\setlength{\epsfxsize}{0.5\linewidth}\leavevmode\epsfbox{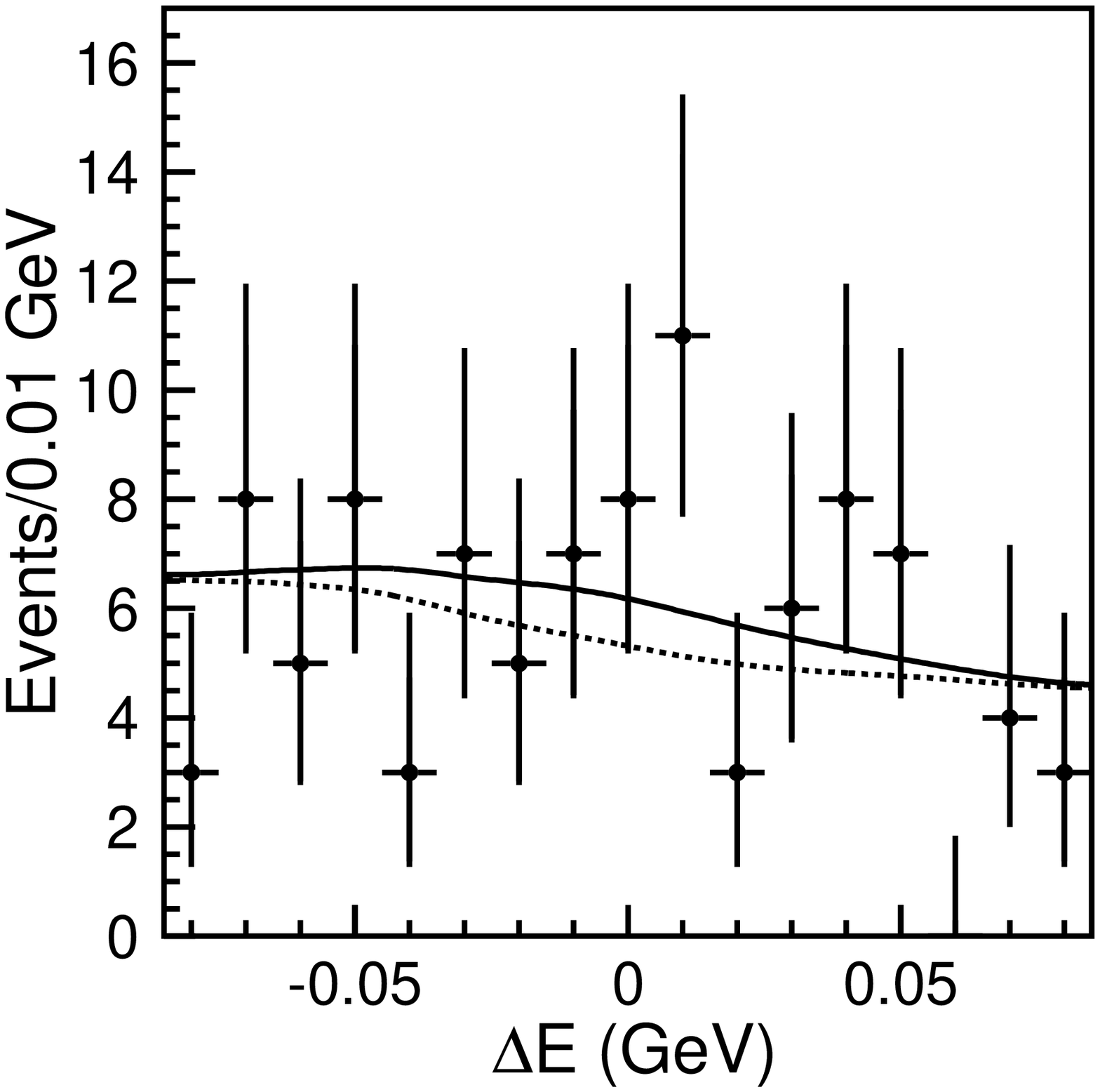}
}
\caption{
Projections of the multidimensional fit onto 
$m_{\rm ES}$ and $\Delta E$ with a requirement on the 
signal-to-background probability ratio
${\cal P}_{\rm sig}/{\cal P}_{\rm bkg}$ with 
the plotted variable excluded. 
The histogram shows the data and the solid (dashed) line 
shows the full (background only) PDF projection.
The projections contain 22.5\% and 23.9\% of signal, and less than
0.5\% and 0.2\% of continuum background, respectively.}
\label{fig:projections}
\end{figure}

Table~\ref{tab:results} shows the results of the fit. No significant
signal yield is observed. 
We obtain an upper limit by integrating
the normalized likelihood distribution over the positive values of 
the branching fraction. 
The value of $f_L$ is fixed to 1 in the fit, as this assumption has 
been shown to give the most conservative upper limit
and it approximates the values obtained in the $B\to\rho\rho$ 
decays dominated by the tree-level amplitude.
The statistical significance is taken as the 
square root of the change in $-2\ln{\cal L}$ 
when the number of signal events is constrained to zero
in the likelihood fit.
In Fig.~\ref{fig:projections} we show the projections of the fit results 
onto $m_{\rm ES}$ and $\DeltaE$.

\begin{table}[b]
  \centering
  \caption{ 
Summary of results: signal yield ($n_{\rm sig}$), selection
efficiency (Eff), branching fraction (${\cal B}$),
branching fraction upper limit (UL) at 90\% CL, 
and significance (including systematic uncertainties).
The assumption $f_L=1$ is used.
The systematic errors are quoted last.
}
\vspace{0.2cm}
  \begin{tabular}{lcc}  
\hline\hline
                     & ~~~~~~~~~ &   \vspace*{-0.3cm} \\
Quantity                      &  &  Value             \\
                              &  &   \vspace*{-0.3cm} \\
\hline                     
                              &  &   \vspace*{-0.3cm} \\
$n_{\rm sig}$ (events)        &  &   $33^{+22}_{-20}\pm 12$ \\
                              &  &   \vspace*{-0.3cm} \\
Eff (\%)                      &  &   $27.1\pm 1.3$       \\
                              &  &   \vspace*{-0.3cm} \\
${\cal B}$ $(\times 10^{-6})$ &  &   $0.54^{+0.36}_{-0.32}\pm 0.19$ \\
                              &  &   \vspace*{-0.3cm} \\
UL $(\times 10^{-6})$         &  &   $1.1$  \\          
                              &  &   \vspace*{-0.3cm} \\
Significance ($\sigma$)       &  &   $1.6$  \\     
                              &  &   \vspace*{-0.4cm} \\
                              &  &   \vspace*{-0.3cm} \\
\hline\hline
  \end{tabular}
  \label{tab:results}
\end{table}

Systematic errors in the fit originate from uncertainties 
in the PDF parameterizations, which arise from the limited 
number of events in the sideband data and signal 
control samples.
The PDF parameters are varied by their respective uncertainties 
to derive the corresponding systematic errors (6.0 events).
The event yields from the $B$-background modes 
fixed in the fit are varied according to the uncertainties 
in the measured or estimated branching fractions. This results 
in a systematic error on the signal yield of 5.8 events.
We also assign a systematic error of 3.0 events to cover 
a possible fit bias, evaluated with MC experiments. 

We estimate the systematic uncertainty due to
signal-$a_1^{\pm}\pi^{\mp}$ interference using a simulation 
study in which the decay amplitudes for $B^0\to\rho^0\rho^0$
are generated according to this measurement 
and those for $B^0\to a_1^{\pm}\pi^{\mp}$ correspond
to a branching fraction of $4\times 10^{-5}$~\cite{a1pi}.
The relative phases between these are modeled with BW amplitudes
for all $\rho\to\pi\pi$ and $a_1\to\rho\pi$ combinations,
with additional constants. The values of the constants and 
the $a_1^{\pm}\pi^{\mp}$ $\CP$ asymmetries
are varied over the allowed ranges.
We take the rms variation of the average signal yield 
(7.5 events) as a systematic uncertainty.

Uncertainties in the reconstruction efficiency
arise from track finding (3\%),
particle identification (2\%),
and other selection requirements, 
such as on vertex probability (2\%), 
track multiplicity (1\%),
and thrust angle (1\%).


Our measurement confirms the small 
value of the $B^0\to\rho^0\rho^0$ branching fraction 
with the statistical uncertainty improved by approximately 
a factor of two over our previous result~\cite{vvbabar}. 
Since the tree contribution to the $B^0\to\rho^0\rho^0$ 
decay is color-suppressed, 
the decay rate is sensitive to the penguin amplitude.
Thus, this mode 
has important implications for constraining 
the uncertainty due to penguin contributions
in the measurement of the CKM unitarity angle $\alpha$
with $B\to\rho\rho$ decays. 

In the isospin analysis~\cite{gronau90}, 
we minimize a $\chi^2$ that includes the measured quantities
expressed as the lengths of the sides of the 
isospin triangles. 
We use the measured branching fractions and 
fractions of longitudinal polarization of the 
$\Bptorhozrrhop$~\cite{vvbabar, rho0rhopbelle}
and $\Bztorhoprhom$~\cite{rhoprhom, rhoprhom2} decays,
the \CP-violating parameters $S^{+-}_{L}$ and $C^{+-}_{L}$
obtained from the time evolution of the longitudinally 
polarized $\Bztorhoprhom$ decay~\cite{rhoprhom2}, and the 
branching fraction of $\Bztorhozrhoz$ from this analysis.
We neglect isospin-breaking effects, non-resonant, 
and $I=1$ isospin contributions~\cite{falketal}.

\begin{figure}[t]
\begin{center}
\setlength{\epsfxsize}{1.0\linewidth}\leavevmode\epsfbox{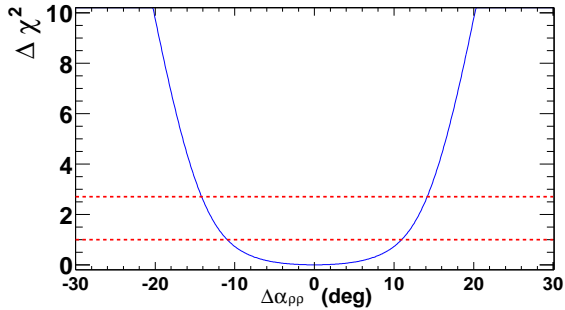}
\caption{$\Delta\chi^2$ on $\Delta\alpha_{\rho\rho}$ 
obtained from the isospin analysis discussed in the text. 
The dashed lines at $\Delta\chi^2 = 1$ and $\Delta\chi^2 = 2.7$ 
are taken for the 1 $\sigma$ (68\%) and 1.64 $\sigma$ (90\%) 
interval estimates.}
\label{fig:alphascan}
\end{center}
\end{figure}

With the \Bztorhozrhoz measurement we improve the
constraint on $\alpha$ due to the penguin contribution 
and obtain a 68\% (90\%) CL limit on 
$\Delta\alpha_{\rho\rho}=\alpha-\alpha_{\rm eff}$ 
of $\pm11^{\mathrm o}$ ($\pm14^{\mathrm o}$).
Fig.~\ref{fig:alphascan} shows the 
$\Delta\chi^2$ on $\Delta\alpha_{\rho\rho}$.
Since the central value from Fig.~\ref{fig:alphascan} 
is $\Delta\alpha_{\rho\rho} = 0$,
the central value of $\alpha$ obtained from the isospin 
analysis is the same as $\alpha_{\rm eff}$, 
which is constrained by the relation
$\sin(2\alpha_{\rm eff})= S^{+-}_{L}/({1-C^{+-2}_{L}})^{1/2}$
and is measured with the $B^0\to\rho^+\rho^-$
decay~\cite{rhoprhom2} to be ${\alpha_{\rm eff} = \left 
( 102^{+16}_{-12} ({\rm stat}) ^{+5}_{-4} ({\rm syst})
\right ) ^{\mathrm o}}$ at 68\% CL,
where the solution closest to the CKM best 
fit central value~\cite{ckm} is chosen.

The error due to the penguin contribution may become
the dominant uncertainty in the measurement of $\alpha$ using
$B\to\rho\rho$ decays. However, if $B^0\to\rho^0\rho^0$ decays are 
observed, time-dependent and angular analyses will allow us 
to measure the \CP\ parameters $S^{00}_{L}$ and $C^{00}_{L}$,
analogous to $S^{+-}_{L}$ and $C^{+-}_{L}$,
resolving ambiguities inherent to isospin triangle orientations.

In summary, we have improved the precision on the measurement 
of the \Btozz branching fraction by approximately a factor of two.
The limit on this branching fraction relative to those for
$B^+\to\rho^0\rho^+$ and $B^0\to\rho^+\rho^-$ provides
a tight constraint on the penguin uncertainty 
in the determination of the CKM unitarity angle $\alpha$.
The results summarized in Table~\ref{tab:results} supersede 
our previous measurement~\cite{vvbabar}.


We are grateful for the excellent luminosity and machine conditions
provided by our \pep2\ colleagues, 
and for the substantial dedicated effort from
the computing organizations that support \babar.
The collaborating institutions wish to thank 
SLAC for its support and kind hospitality. 
This work is supported by
DOE
and NSF (USA),
NSERC (Canada),
IHEP (China),
CEA and
CNRS-IN2P3
(France),
BMBF and DFG
(Germany),
INFN (Italy),
FOM (The Netherlands),
NFR (Norway),
MIST (Russia), and
PPARC (United Kingdom). 
Individuals have received support from CONACyT (Mexico), A.~P.~Sloan Foundation, 
Research Corporation,
and Alexander von Humboldt Foundation.


\bibliographystyle{h-physrev2-original}  

\begin{thebibliography}{99}

\bibitem{CabibboKobayashi}
M. Kobayashi, T. Maskawa, Prog. Theor. Phys. {\bf 49}, 652 (1973);
N. Cabibbo, Phys. Rev. Lett. {\bf 10}, 531 (1963).

\bibitem{gronau90}
M.~Gronau, D.~London, \jprl{65}, 3381 (1990).

\bibitem{pi0pi0}
\babar\ Collaboration, B.~Aubert {\it et al.}, 
hep-ex/0412037, submitted to \jprl

\bibitem{vvbabar}
\babar\ Collaboration, B.~Aubert {\it et al.},
\jprl{91}, 171802 (2003).

\bibitem{rho0rhopbelle}
Belle Collaboration, J.~Zhang {\it et al.},
\jprl{91}, 221801 (2003).

\bibitem{rhoprhom}
\babar\ Collaboration, B.~Aubert {\it et al.},
Phys. Rev. D {\bf 69}, 031102 (2004).

\bibitem{rhoprhom2}
\babar\ Collaboration, B.~Aubert {\it et al.},
\jprl{93}, 231801 (2004).

\bibitem{babar}
\babar\ Collaboration, B.~Aubert {\it et al.},
Nucl. Instrum. Methods Phys. Res., 
Sect. A {\bf 479}, 1 (2002).

\bibitem{pep2}
PEP-II Conceptual Design Report, SLAC-R-418 (1993).

\bibitem{bigPRD}
$\babar$ Collaboration, B.~Aubert {\it et al.},
\jprl{89}, 281802 (2002);
Phys.\ Rev.\ D {\bf 70}, 032006 (2004).

\bibitem{GEANT}
The \babar\ detector Monte Carlo simulation is based
on GEANT4: S. Agostinelli {\it et al.},
Nucl. Instrum. Methods Phys. Res., 
Sect. A {\bf 506}, 250 (2003).

\bibitem{babarsin2beta}
\babar\ Collaboration, B.~Aubert {\it et al.},
\jprl{89}, 201802 (2003).

\bibitem{argus}
ARGUS Collaboration, H.~Albrecht {\it et al.},
Z.\ Phys.\ C {\bf 48}, 543 (1990).

\bibitem{nonparam}
K.S.~Cranmer, Comput.\ Phys.\ Commun. {\bf 136}, 198 (2001).

\bibitem{a1pi}
\babar\ Collaboration, B.~Aubert {\it et al.}, 
hep-ex/0408021.

\bibitem{falketal}
A.F.~Falk {\it et al.}, Phys.\ Rev.\ D {\bf 69}, 011502 (2004).

\bibitem{ckm}
J.~Charles {\it et al.}, hep-ph/0406184, submitted to Eur. Phys. J. C;
M.~Bona {\it et al.}, hep-ph/0408079.

\end{thebibliography}

\end{document}